\documentclass{article}

\usepackage[final,nonatbib]{nips_2017}

\usepackage[utf8]{inputenc} 
\usepackage[T1]{fontenc}    
\usepackage[draft]{hyperref}       
\usepackage{url}            
\usepackage{booktabs}       
\usepackage{amsfonts}       
\usepackage{nicefrac}       
\usepackage{graphicx,float}
\usepackage{amssymb,amsmath,amsthm}
\usepackage{cleveref}
\usepackage[square,numbers]{natbib}

\usepackage{color}
\definecolor{lightgray}{gray}{0.85}

\newcommand\greybox[1]{%
  \vskip\baselineskip%
  \par\noindent\colorbox{lightgray}{%
    \begin{minipage}{0.95\textwidth}#1\end{minipage}%
  }%
  \vskip\baselineskip%
}

\newcommand{\ignore}[1]{}

\usepackage{color}

\newcommand{\coin}{PROBO\,}
\newcommand{\coins}{\emph{probos}\,}

\title{Towards a scientific blockchain framework for reproducible data analysis}

\author{
Cesare Furlanello${}^{1}$\thanks{Corresponding author} $\quad$ Manlio De Domenico${}^{2}$\thanks{Corresponding author; CF and MDD joint first author} $\quad$ Giuseppe Jurman${}^{1}$ $\quad$ Nicole Bussola${}^{1}$\\
${}^{1}$ Fondazione Bruno Kessler, Trento, Italy\\
${}^{2}$ Universitat Rovira i Virgili, Tarragona, Spain\\
\texttt{furlan@fbk.eu, manlio.dedomenico@urv.cat, jurman@fbk.eu, bussola@fbk.eu}
}

\begin{document}
\maketitle

\begin{abstract}
Publishing reproducible analyses is a long-standing and widespread challenge~\citep{lawrence1989concordance} for the scientific community, funding bodies and publishers~\citep{collins2014nih,nature2017reproducibility,munafo2017manifesto}. Although a definitive solution is still elusive~\citep{baker20161}, the problem is recognized to affect all disciplines~\citep{prinz2011believe,franco2014publication,open2015estimating} and lead to a critical system inefficiency~\citep{gilbert2016comment}. Here, we propose a blockchain-based approach to enhance scientific reproducibility, with a focus on life science studies and precision medicine. While the interest of encoding permanently into an immutable ledger all the study key information--including endpoints, data and metadata, protocols, analytical methods and all findings--has been already highlighted, here we apply the blockchain approach to solve the issue of rewarding time and expertise of scientists that commit to verify reproducibility. Our mechanism builds a trustless ecosystem of researchers, funding bodies and publishers cooperating to guarantee digital and permanent access to information and reproducible results. As a natural byproduct, a procedure to quantify scientists' and institutions' reputation for ranking purposes is obtained. 
\end{abstract}

\section{Introduction}
\label{sec:intro}
Reproducibility is a key property of scientific endeavour~\citep{goodman16what}, enabling the progressive structuring of knowledge into innovation through a process of knowledge integration and reuse in downstream studies. The continuous sharing of assets such as information, data and tools implies incentives, trust and accountability within an ecosystem of networks. In life science we can count at least three different networks, each mostly formed by competing actors: academic research, pharmaceutical industry and funding bodies. Pharma companies systematically attempt to replicate academic studies with a potential for clinical application: however, the total prevalence of irreproducible preclinical studies exceeds 50\%, each replication requiring between \mbox{US\$500,000} to \mbox{US\$2,000,000} investment \citep{PhRMA2013}. For 2012, a prevalence of irreproducible preclinical research was estimated to exceed 50\%, resulting in approximately US\$28B/year spent in the United States alone according to \cite{freedman2015economics}. Thus, a low rate of reproducibility within life science research not only limits knowledge integration and reuse, but it also ramps up costs of developing therapeutic drugs and it delays benefit to patients. 

In order to accelerate the discovery of life-saving therapies, the key to a solid improvement in reproducibility rate is to identify the critical leverage points inducing a change in the ecosystem; \cite{rosenblatt2016incentive} advocated that investigators and their institutions should receive specific financial incentives to diminish the data irreproducibility problem at the academia-industry interface, but these should be matched by a form of money-back guarantee. However, \cite{topol2016moneybg} argued that reproducing findings is not a simple story and it requires time and experience; further, even clinical studies cannot be trusted as immune of irreproducibility in terms of clinical efficacy and risk. Both in academia and life science industry, the first obstacle to remove is the lack of verified compliance to a fully transparent access to research findings, data and protocols, including the capability of tracking of all decisions and changes along the whole course of experiments. This is a lesson that machine learning scientists as well as genomics experts have already learnt the hard way. Developing predictive models from massive genomics data carries a pervasive risk of selection bias \citep{Ambroise2002,furlanello2003entropy} and it requires a careful evaluation of all causes of variability from bioinformatics procedures to biomarker identification \citep{maqc10maqcII}, and overall a deep commitment to provide access to data, metadata and protocols. \cite{ioannidis2009repeatability} found that almost 90\% of microarray studies from a leading genetics journal was found not fully repeatable, mostly due to incomplete availability of data and methods. Even in the case of full availability of data, software and methods, it has been recently pointed out that peer review and editor evaluation provide just a filter to poor data analysis that, in more complex studies, needs preventive measures such as education to data science and evidence-based data analysis~\citep{leek2015opinion}.

The blockchain is a decentralized and distributed technology~\citep{bitcoinwhite} developed to avoid the necessity of authorities that regulate the exchange of money and any type of assets~\citep{ethwhite}, from music to health care. The blockchain relies on strong cryptographic protocols which are able to guarantee the secure, (pseudo-)anonymous and trustless interactions among peers of the underlying network. The transaction, \emph{e.g.} a money transfer, is independently verified by several \emph{miners} -- nodes of this system with high computational power which own a copy of the distributed ledger of any previous transaction -- who solve complex mathematical puzzles and have to reach consensus about its validity. If consensus is achieved, the transaction is accepted and encoded into a block of data which is permanently stored in the blockchain.

The blockchain technology can be key to address the issues of replicability, accountability and trust in scientific studies, providing an immutable ledger for all the steps, from protocols to all outcomes. One first implementation was sketched in \cite{carlisle2014timestamp}, as a blockchain-based solution to notarize univocally the endpoint and the analytical strategies into the permanent, timestamped Bitcoin public ledger. Since then, the enthusiasm for blockchain-based solutions has raised so high to be considered indispensable to the future of precision medicine ecosystem, improving data security, data sharing, interoperability, patient engagement by wearable sensors, AI-based diagnostics and much more \citep{shumacher2017}. 

As discussed, all members of the scientific ecosystem---from researchers to funding bodies and publishers---are interested in fostering reproducibility to save time and investments that would otherwise fund bad practices and misleading outcomes. However, the current system does not reward appropriately those members of the community who invest in conducting rigorous and reproducible research~\citep{munafo2017manifesto}. Here we fill this gap by proposing a blockchain-based mechanism to ignite the interest in reproducing scientific analyses by rewarding time and expertise of researchers and institutions that commit to verify reproducibility. Our mechanism is trustless, transparent, resilient to scientific fraud and traceable, providing as a byproduct a natural procedure to rank authors and institutions for their contribution to advancing science with reproducible results.

The current science ecosystem -- including interdependent networks of researchers, institutions and funding bodies -- is more disintegrated than ever by competition for limited resources, with key players in a constant rush for publishing faster than others in the best journals and trying to centralize the destination of available funding. Our mechanism, that we name \coin network in the following, aims at restoring cooperation among the main players of this ecosystem, by providing a natural framework to achieve those ideals of transparency, reproducibility and credibility that should characterize science and scientific knowledge.

\begin{figure}
\centering
\includegraphics[width=14cm]{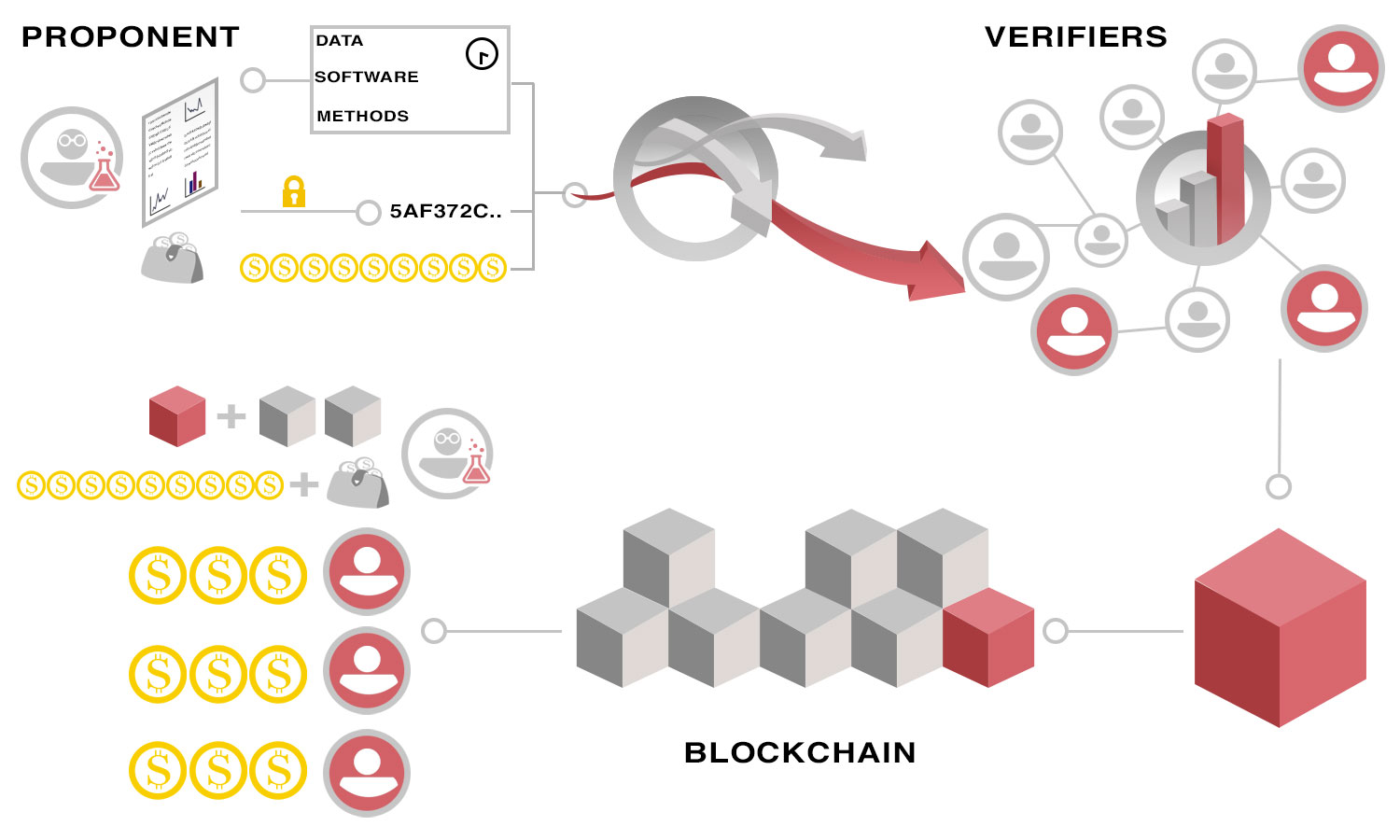}
\caption{\label{fig1}\textbf{Illustration of the PROBO network.} A researcher/lab (the \emph{proponent}) aims to publish her last study in the public ledger. Access to the collection of data, software and methods is provided as input, while the results are encrypted and safely encoded by a hash function. The proponent deposits a pre-determined amount of \coins coins to start a smart-contract-like mechanism which broadcasts her request to the PROBO network, where other researchers/labs (the \emph{verifiers)} will evaluate the quality of the study and verify its reproducibility. If enough verifiers agree on the study reproducibility, a block encoding all relevant information is generated and added to the existing PROBO blockchain. The system generates new \coins to reward the verifiers' efforts, as well as to refund their expenses, and transfers back the deposit to the proponent---thus finalizing the smart contract---who collects one more block for her scientific \emph{curriculum}. See the text for the case of rejected requests, and other details.
}
\end{figure}

\section{A reproducibility mechanism: the \coin blockchain}

To cope with the issue of data analysis reproducibility, first we assemble \coin as a sustainable blockchain ecosystem: this requires defining the blockchain \emph{ledger} where the results of an elementary transaction (a study to verify) will be notarized and a set of rules for rewarding consistently with \emph{coins} (tokens) the \emph{mining} agents entitled to verification. We aim to adapt known blockchain concepts to incentive peers' evaluation, thus we focus more on the critical mechanisms rather than on implementing a set of encoding hacks. The first aim is to create a self-supporting network including all actors. The nodes of the \coin network build an ecosystem of well established researchers, research institutions and funding bodies, identified by a unique and persistent digital identifier such as the ORCID. Different approaches can be adopted to ignite the system, from sharing a predetermined amount of tokens like by an initial coin offering (ICO) to equally distributing a fixed amount of digital currency. A technical description of this phase is beyond the scope of the present work and requires iterative interactions among all parties of the ecosystem to reach an agreement.


The proposed mechanism is summarized and illustrated in Fig.\,\ref{fig1}. One or more nodes of the network produce a scientific work based on a data analysis, which is required to be reproduced by independent peers. Note that we use the word reproducibility in its broader meaning, including obtaining results that are not duplicating the original study but they are consistent with them, \textit{i.e.} replicability. To this aim, the \emph{proponent} prepares the inputs as a time-stamped collection of data, software and methods used for the analysis, providing direct and persistent access to this immutable resource (\textit{e.g.}, by using existing services such as Github, Bitbucket or FigShare). In parallel, the proponent prepares a standardized collection of outputs---encoding by tables the results of the data analysis----and computes an hash by means of complex cryptographic algorithms, to generate a long string of characters which uniquely identifies the outputs. The proponent also sets the expected acceptable reproducibility boundaries of their study by providing a set of statistical indicators in a standardized format, including false positives rate (Type I error) and false negatives rate (Type II error), as well as true positives and negatives rates to allow for the comparison of outputs while accounting for variation across laboratories, systematic biases and error propagation. Recent studies have shown that this goal can be achieved by appropriately distributing tasks with proper standardization and cross-check procedures~\citep{hoen2013reproducibility}. 
The proponent shall specify the expected amount of time required to replicate the study, within a time horizon set by the network.

Finally, the proponent creates a block request, including the inputs and the hash of the outputs, which is broadcasted to the \coin network. It is worth remarking that such an action might also encode a publication request to a publisher, provided that the publisher is part of this ecosystem. To prevent nodes from flooding the system with research that is not reproducible or that is not worth reproducing, a smart-contract-like mechanism is introduced~\citep{ethwhite}. In fact, a certain amount of \coins---the digital currency of the system, which can be used to fund further research, buy equipment, and so on---is asked to be deposited from the proponent for allowing their request to be broadcasted to the network.
It is worth accounting for scenarios when one proponent requires a pipeline or a set of complementary experiments to be written in the blockchain. In that case, it is expected to broadcast each experiment separately to facilitate the reproducibility of each phase of the overall experiment.

The final request is transformed into a pending transaction that is freely accessible from any other node of the network different from the proponent. Nodes, that we might indicate as \emph{verifiers}, decide which pending request is worth accepting, according to their assessment of the quality of the research and their skills (\textit{e.g.}, it is very unlikely that a node working mainly in translational medicine will attempt to reproduce a study from quantitative psychology). Each verifier uses the same data, (version of) software and methods of the proponent to produce an hash of the standardized outputs. Finally, the verifier compares her hash with the one of the proponent and if they match (mismatch), the verifier accepts (rejects) the transaction. 

A minimum number of verifiers is required in order to definitively accept or reject the proponent request. In case of acceptance, the system generates a block of information including the inputs, the outputs, the hash of the outputs, the identity of proponent and verifiers, as well as the time-stamp of each activity. The block is therefore added to the blockchain, the system sends back the deposited \coins to the proponent and it generates a predefined number of \coins to reward the efforts and the expenses of the verifiers. Conversely, in case of rejection, the system does not generate new \coins and it rewards the verifiers by transferring the \coins deposited by the proponent. As for the case of acceptance, the rejection details are encoded into a block which is added to the blockchain, to keep trace of the proponent's attempt.

The PROBO ecosystem should also be effective in uncovering scientific misconduct in its early stage, thus nipping in the bud unwanted horror stories running through several years (e.g. see \citep{reich2009plastic}), and made possible by an outrageous lack of reproducibility throughout all the steps of the research.

\section{Discussion}

The blockchain-based approach that we envision is expected to mitigate some of the issues which are currently affecting the reproducibility of scientific analyses by widening the incentive offer. Its sustainability will depend on both technical and social mechanisms that need to be acknowledged and embedded in the design of the PROBO network. Here we sketch a list of relevant aspects, knowing that additional elements may need to be detailed.

The total supply of \coins in the \coin network will depend on multiple factors, as for other existing blockchains. For instance, the total supply of bitcoins is limited to 21 million by the integer storage type of the transaction output, which can be easily extended in new-generation blockchains like the one proposed here. The total supply will be mostly affected by the parameters regulating verifiers' recompense -- like the ratio between proponent's deposit and block reward -- in the case of study acceptance, which will be determined in order to optimize the efficiency of the \coin network and the benefits of its nodes.

Besides the stimulus triggered by the \coin network on activities dedicated to verify reproducibility, an interesting byproduct is that this blockchain allows a quantitative evaluation of scientists' and institutions' prestige. In fact, the number of blocks written in the blockchain by proponents and verifiers can be used to build the reputation of researchers -- and, at a coarser level, of institutions -- which in turn provides a direct proxy for the quality of their research and, overall, the reproducibility of their analyses. Failures, \emph{i.e.} transaction requests that have not been accepted by the network for lack of reproducibility, are also accounted for, pushing researchers and institutions to be more careful with the analyses they intend to add in the blockchain.

Therefore, the proposed blockchain-like mechanism embeds a direct, transparent, objective and unbiased way for funding bodies to rank scientists and institutions. This reputation score is expected to be a valid alternative to Hirsch index~\citep{hirsch2005index} and other bibliometrics measures widely adopted for ranking with respect to a variety of criteria, from performance to interdisciplinarity~\citep{radicchi2008universality,bollen2009principal,radicchi2009diffusion,nature2010metrics,wang2013quantifying,sinatra2016quantifying,omodei2017evaluating}.

Additionally, ranking by the reputation score can be accounted for by decision makers to hire or fund researchers in a trustless ecosystem; indeed the quality of scientific content and its reproducibility should drive decisions without relying -- only -- on bibliometrics data, which can be mesleading or distorted, or even fake  ~\citep{folly1981some,van2005fatal,jalalian2015fake}. Further, the same ranking score can be adopted by journals as an additional dimension of their own prestige in terms of reproducibility and good scientific practice, based on the papers either published or rejected, whose studies were submitted to the \coin blockchain. 

This architecture may attract big research institutions or labs to investing into ``reproducibility units'', new groups of researchers working as verifiers in the \coin network to produce a steady income in terms of \coins to fund further research. Indeed a first attempt to improve the quality of biomedical literature as the Reproducibility Initiative, fostering independent corroboration of published results~\citep{reproinit2012}, was initially supported by scientist, publishers, and private foundations and it is now being implemented through a online marketplace for scientific experiments. Of course there is a risk that without an embedded feedback mechanism towards authors or institutions' reputation, the birth of scientific facilities specialized in testing reproducibility for third parties will have no effect or even ironically sustain the current crisis.  

It is worth remarking that, as a community-based effort, the laboratories' commitment to our blockchain mechanism relies on the same philosophy that inspired the quest for standards and open access to material and methods in computational biology (and -omics studies in particular) throughout the last twenty years. Starting with the MIAME standard for microarray experiments proposed by the FGED Society  ~\citep{brazma2001minimum}, a number of similar protocols for the omics types has led first to a rationalisation attempt~\citep{taylor2008promoting} and then to the comprehensive FAIR initiative~\citep{wilkinson2016fair} where ``a diverse set of stakeholders---representing academia, industry, funding agencies, and scholarly publishers---have come together to design and jointly endorse a concise and measurable set of principles that we refer to as the FAIR Data Principles''. The FAIR (Findability, Accessibility, Interoperability, and Reusability) guidelines are aimed at maximizing the added-value attached to formal scholarly digital publishing in the broader sense, thus including data, algorithms, tools, and workflows in order to warrant transparency, reproducibility, and reusability. All these requirements would facilitate the procedural description of an analysis to replicate (see example in Box 1). 

In summary, the blockchain-based network like the one we design in this paper has naturally built in a rich set of options for explicit and implicit feedback mechanisms based on standardized description of data, methods and outcomes. We argue that the ecosystem spanned by the \coin network may intercept a wider set of incentives than existing practice aiming to obtain a pervasive adoption of open science and reproducibility concepts. 

\begin{figure}[!t]
\greybox{
\textbf{Box 1: Example of Bioinformatics Use Case with PROBO}
\vskip0.2cm
\scriptsize
The pipeline defined in this Descriptor implements an integrative classification task in computational biology, based on multitype data and a sequential combination of methods from Bioinformatics and Machine Learning. For brevity, exact specification of all the involved software packages is omitted; however it is intended that each item should be indicated with reference to a persistent resource, specifying versions and all relevant features.   
\begin{description}
\item[Aim] Discriminating a cohort of SCCLC  patients from a control set of healthy subjects, identifying a predictive classifier model and top biomarkers of interest.
\item[Data] For each patient: collection of chest CT scans + RNA-Seq transcriptomic profile of lung tumoral (resp. healthy) tissues (for the patients) or of normal lung tissue (for the controls). 
\item[Preprocessing] 3D images are segmented, converted to a file format suitable as classifier input and reshaped to same size; RNA-Seq data preprocessing pipeline includes quality control, read alignment, batch effect reduction, transcript identification and quantification \cite{conesa2016survey}
\item[Analysis] A 10 x 5 Cross Validation Data Analysis Protocol (DAP) \citep{maqc10maqcII} with data splitting and resampling (to avoid overfitting effects) with a Deep Neural network as classifier/feature ranker.
\item[Outcome] A summary table including different statistical indicators of classifier performance (\textit{e.g.}, MCC) with corresponding confidence intervals + a ranked list of top discriminating features.
\end{description}
\vskip0.2cm
\textbf{\coin Proponent input block}
\begin{description}
\item{Data \& Metadata}
\begin{itemize}
\item Univocal name and version of the CT scan hardware, with the description of the raw image format
\item Link to the web resource where each raw image can be accessed with its hash (\textit{e.g.} md5) 
\item A text string linking the raw image filename with the matching patient/control
\item an univocal name and version of the sequencing device with the description of the read file format, in MAGE-TAB standard 
\item Link of the web resource where each file can be accessed with its hash
\item a text string linking the raw data filename with the matching patient/control case
\end{itemize}
\item{Preprocessing}
\begin{itemize}
\item Name and version of preprocessing or formatting (e.g. to unique window shape size), with parameters used for each image if needed
\item Link to the web resource where each preprocessed image can be accessed with its hash 
\item A text string linking the preprocessed image filename with the matching patient/control
\item Name and version of the RNA-Seq read alignment software; the genome version and the parameters used in running the read alignment; the name and version of the transcript quantification software; the parameters used in running the quantification; information about sample batches, used for batch correction; the name and version of the batch correction software.
\item the ASCII version of the samples $\times$ features matrix including all preprocessed RNA-Seq data
\end{itemize}
\item{Analysis}
This section may be summarized by a notebook including all step-by-step details
\begin{itemize}
\item An unambiguous description of the Data Analysis Protocol with the involved parameters							\item name/version of the software used for implementing the DAP
\item A working software implementation of the analysis pipeline (all subcomponents)
\item Hardware characteristics and configuration of the machine(s) running the code
\item (Deep Learning specific) description of the architecture, including layers of the neural networks with the corresponding parameters (batch size, number of epochs, dropout, etc.)
\item Name/version of the software used for the neural network implementation
\end{itemize}
\end{description}
\textbf{\coin Proponent output block}
The hash of the summary table, including a ranked list of features, confidence interval for expected acceptable outcomes and expected max time to validate (ETV). 
}
\end{figure}





\subsubsection*{Acknowledgments}
The authors wish to thank Serafina Agnello for graphics, Marco Chierici for details on the Next Generation Sequencing bioinformatics workflow, and Marco Ajelli for comments. MDD acknowledges financial support from the MINECO (Spain) program ``Juan de la Cierva'' (IJCI-2014-20225) and the kind hospitality of Fondazione Bruno Kessler.

\bibliographystyle{unsrtnat}
\small
\bibliography{main}
\end{document}